\documentclass[conference]{IEEEtran}
\ifCLASSINFOpdf
\else
\fi

\usepackage{float}
\usepackage{graphicx}
\usepackage{color}
\usepackage[square, comma, sort&compress, numbers]{natbib}%\ggg
\usepackage{subfigure}
\usepackage{amsmath}
\usepackage{algorithm}%\hhh
\usepackage{algorithmic}%\hhhh
\usepackage{amssymb}%\hhhh
\usepackage{threeparttable}
\usepackage{booktabs}
\usepackage{makecell}
\usepackage{colortbl}
\usepackage{multirow}
\usepackage{amsfonts}
\usepackage{amssymb}
\usepackage[draft]{hyperref}
\usepackage{url}

\usepackage{graphicx}
\definecolor{mygray}{gray}{.9}
\begin{document}
% paper title
% Titles are generally capitalized except for words such as a, an, and, as,
% at, but, by, for, in, nor, of, on, or, the, to and up, which are usually
% not capitalized unless they are the first or last word of the title.
% Linebreaks \\ can be used within to get better formatting as desired.
% Do not put math or special symbols in the title.
\title{Molecular Polar Belief Propagation Decoder and Successive Cancellation Decoder}
% autho names and affiliations
% use a multiple column layout for up to three different
% affiliations
\author{\IEEEauthorblockN{Zhiwei Zhong$^{1,2,3}$, Lulu Ge$^{1,2,3}$, Zaichen Zhang$^{2,3}$, Xiaohu You$^{2}$, and Chuan Zhang$^{1,2,3,*}$}
\IEEEauthorblockA{
$^1$Lab of Efficient Architectures for Digital-communication and Signal-processing (LEADS)\\
$^2$National Mobile Communications Research Laboratory, Southeast University, Nanjing, China\\
$^3$Quantum Information Center of Southeast University\\
%$^{1}$Lab of Efficient Architectures for Digital-communication and Signal-processing (LEADS)\\
%$^{2}$National Mobile Communications Research Laboratory, Southeast University, Nanjing, China\\
%%$^{3}$State Key Laboratory of Coordination Chemistry,
%%%School of Chemistry and Chemical Engineering,
%%Nanjing University, Nanjing, China\\
Email: \{zwzhong, chzhang\}@seu.edu.cn}

%{\em (Invited Paper)}
}
% conference papers do not typically use \thanks and this command
% is locked out in conference mode. If really needed, such as for
% the acknowledgment of grants, issue a \IEEEoverridecommandlockouts
% after \documentclass

% for over three affiliations, or if they all won't fit within the width
% of the page, use this alternative format:
%
%\author{\IEEEauthorblockN{Michael Shell\IEEEauthorrefmark{1},
%Homer Simpson\IEEEauthorrefmark{2},
%James Kirk\IEEEauthorrefmark{3},
%Montgomery Scott\IEEEauthorrefmark{3} and
%Eldon Tyrell\IEEEauthorrefmark{4}}
%\IEEEauthorblockA{\IEEEauthorrefmark{1}School of Electrical and Computer Engineering\\
%Georgia Institute of Technology,
%Atlanta, Georgia 30332--0250\\ Email: see http://www.michaelshell.org/contact.html}
%\IEEEauthorblockA{\IEEEauthorrefmark{2}Twentieth Century Fox, Springfield, USA\\
%Email: homer@thesimpsons.com}
%\IEEEauthorblockA{\IEEEauthorrefmark{3}Starfleet Academy, San Francisco, California 96678-2391\\
%Telephone: (800) 555--1212, Fax: (888) 555--1212}
%\IEEEauthorblockA{\IEEEauthorrefmark{4}Tyrell Inc., 123 Replicant Street, Los Angeles, California 90210--4321}}

% use for special paper notices
%\IEEEspecialpapernotice{(Invited Paper)}
% make the title area
\maketitle
% As a general rule, do not put math, special symbols or citations
% in the abstract
\begin{abstract}
By constructing chemical reaction networks (CRNs), this paper proposes a method of synthesizing polar decoder using belief propagation (BP) algorithm and successive cancellation (SC) algorithm, respectively. Theoretical analysis and simulation results have validated the feasibility of the method.
Reactions in the proposed design could be experimentally implemented with DNA strand displacement reactions, making the proposed polar decoders promising for wide application in nanoscale devices.
\end{abstract}
% no keywords
\begin{IEEEkeywords}
Molecular computation, CRNs, polar codes, belief propagation (BP), successive cancellation (SC).
\end{IEEEkeywords}
\IEEEpeerreviewmaketitle
\section{Introduction}
Molecular computation has been widely used as an effective way to carry out probabilistic calculation by regarding the concentrations of molecules as probabilities. \cite{salehi2015markov}  implemented the first-order Markov chain with molecular reactions to solve \emph{Gambler's Ruin Problem}. \cite{shen2016synthesis} employed molecular reactions to solve law-of-total-probability-relevant problems within the realm of probability theory. Given that the essence of decoding channel codes is the computation of probabilities, \cite{xing2018dna} proposed the molecular implementation of belief propagation (BP) decoder for low-density parity-check (LDPC) codes.

Generalizing this idea, we propose the method of synthesizing polar BP decoder and successive cancellation (SC) decoder by constructing relevant chemical reaction networks (CRNs). In the design approach, the decoding formulas, feedback parts of SC decoder, and estimation of source word are implemented by molecular reactions, respectively. Fuel molecules are also introduced to control and activate decoders. Compared with \cite{xing2018dna}, we use fewer reactions for the decoding formulas and only employ catalytic reactions (e.g., $A + B \xrightarrow[]{k} C + D$) in our design. Consider that \cite{chen2013programmable} has preformed DNA experiments for this type of catalytic reaction and validated its robustness and feasibility in $vitro$, we believe that the proposed molecular polar BP decoder and SC decoder could be achieved in nanoscale devices in the future. Note that we propose the CRN-based decoders for polar codes to demonstrate its feasibility and potential use in biocommunication instead of improving the decoding performance. Therefore, the CRNs in this paper conform to conventional BP and SC algorithms and are expected to achieve the same function. We demonstrate the robustness of the CRNs by solving \textit{ordinary differential equations} (ODEs) based on mass-action kinetics \cite{strogatz2018nonlinear}.

The remainder of this paper is organized as follows. Section \ref{sec:1} gives out the preliminaries of polar codes. Section \ref{sec:2} proposes the molecular implementation of formulas used in polar decoding. Section \ref{sec:3} gives out the molecular implementation method and simulation results of BP decoder and SC decoder.  Section \ref{sec:4} remarks on the entire paper.

% You must have at least 2 lines in the paragraph with the drop letter
% (should never be an issue)
\section{Preliminaries}\label{sec:1}
\subsection{Polar Codes}
Polar code \cite{arikan2009channel} is the first channel code that can achieve the capacity of the binary-input discrete memoryless channels (BDMCs). Polar code is represented by a parameter vector of ($N,K,\mathcal{A},u_{\mathcal{A}^c}$), where $N$ is the code length, $K$ is the number of information bits, $\mathcal{A}$ is the set of information bits, and $u_{\mathcal{A}^c}$ is set of frozen bits whose values are fixed to zero.

The construction of polar codes consists of two steps. The first step is to assign the information bits into the $k$ best positions to form the source word $u_1^N$. The second step is to multiply $u_1^N$ and an $N$-by-$N$ generation matrix ${\bf{G}}_N$ to finally obtain the codeword $x_1^N = u_1^N {\bf{G}}_N$. After the transmission of the codeword $x_1^N$, a corrupted codeword $y_1^N$ is received at the receiver. Polar decoder is employed to recover the source codeword $u_1^N$ from $y_1^N$. Polar codes could be decoded with two types of decoding algorithms, namely BP algorithm and SC algorithm. Readers could refer to \cite{arikan2009channel} for more details.

\subsection{Polar BP Decoder}
\begin{figure}[ht]
\centering
\includegraphics[width=6cm]{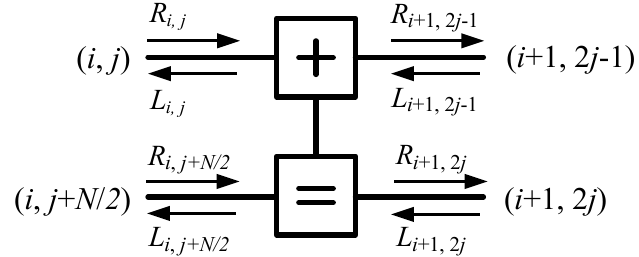}
\caption{The basic computation block in BP decoder.}\label{fig:n2}
\end{figure}

For a ($N,K,\mathcal{A},u_{\mathcal{A}^c}$) polar code, BP decoding process could be represented by a factor graph consisting of $n=\log_{2}N$ stages and $(n+1)N$ nodes. Each factor graph is composed of basic computation blocks (BCBs) in Fig. \ref{fig:n2}.

\begin{figure}[ht]
\centering
\includegraphics[width=7cm]{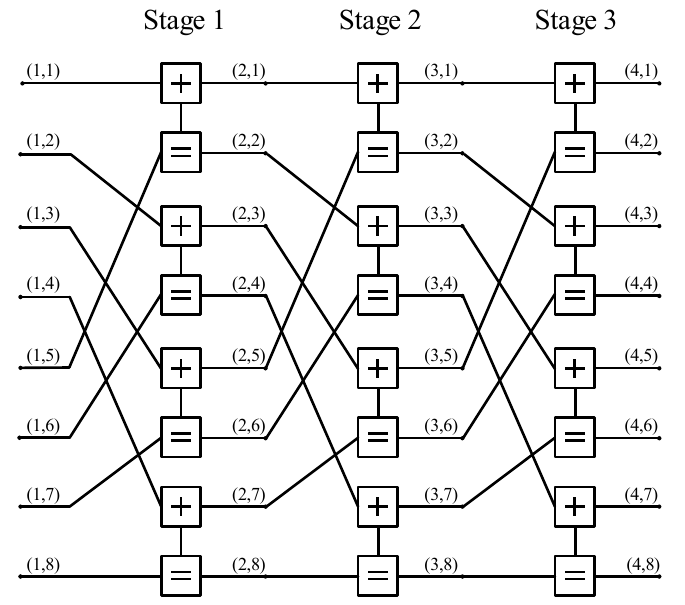}
\caption{The factor graph of polar codes with $N=8$.}\label{fig:n1}
\end{figure}

Fig. \ref{fig:n1} illustrates the factor graph for polar codes with $N=8$. Each node labelled with $(i,j)$ has two types of likelihood ratio (LR) messages, namely left-to-right message $L_{i,j}$ and right-to-left message $R_{i,j}$. Messages $L$ of the right-most nodes and messages $R$ of the left-most nodes of the factor graph are initialised by Eq. (\ref{eq:n0}) and Eq. (\ref{eq:n0.5}), respectively. Values of the rest messages are initialized by zero. Note $L(n+1,j)$ ($j=1,...,N$) is the output vector from the channel whose conditional probability is $P(y_i | x_i)$.
\begin{equation}\label{eq:n0}
    \begin{array}{ll}
     L_{n+1,j}= \dfrac{P(y_j | x_j = 0 )}{P(y_j | x_j = 1 )};
	\end{array}
\end{equation}
\begin{equation}\label{eq:n0.5}
	R_{1,j}  = \begin{cases}
	1, & \text{if $j \in \mathcal{A}$,}  \\
	+\infty, & \text{if $j \in \mathcal{A}^c$,}
	\end{cases}
\end{equation}

Messages $L$ and $R$ are updated and passed iteratively in accordance with:
\begin{equation}\label{eq:n1}
  \left\{
    \begin{array}{ll}
      L_{i,j}=f(L_{i+1,2j-1},g(L_{i+1,2j},R_{i,j+N/2})),\\
	  L_{i,j+N/2}=g(f(R_{i,j},L_{i+1,2j-1}),L_{i+1,2j}),\\
	  R_{i+1,2j-1}=f(R_{i,j},g(L_{i+1,2j},R_{i,j+N/2})),\\
      R_{i+1,2j}=g(f(R_{i,j},L_{i+1,2j-1}),R_{i,j+N/2}),
	\end{array}
  \right.
\end{equation}
where
\begin{equation}\label{eq:n2}
    \begin{array}{ll}
     f(x,y)=\dfrac{1 + xy}{x+y}, \ \ \ \   g(x,y)=xy.
	\end{array}
\end{equation}

During BP processing, the BCBs in the same stage are activated simultaneously. After the maximum number of iteration is achieved, the source word is estimated by:
\begin{equation}\label{eq:n4}
    \begin{array}{ll}
    \hat{u}_j=\left\{
\begin{aligned}
0, &\ \ \quad \text{if} \ L_{1,j} \geqslant 1 \ \ \text{or} \ \ j \in \mathcal{A}^c  \\
1, &\ \ \quad \text{if} \  L_{1,j} < 1  \ \ \text{and} \ \ j \in \mathcal{A}
\end{aligned}
\right.
	\end{array}
\end{equation}

\subsection{Polar SC Decoder}
\begin{figure}[ht]
\centering
\includegraphics[width=7cm]{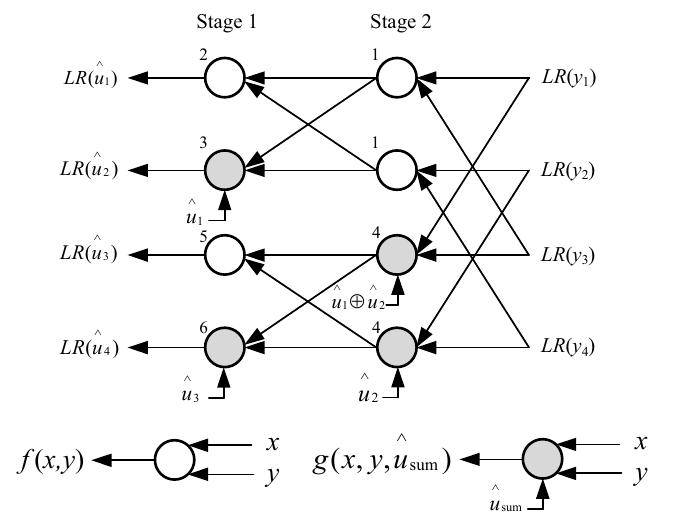}
\caption{SC decoder of polar code with $N=4$ .}\label{fig:n3}
\end{figure}

The SC decoder for a ($N,K,\mathcal{A},u_{\mathcal{A}^c}$) polar code is represented by a graph with $n=\log_{2}N$ stages. Fig. \ref{fig:n3} illustrates the SC decoding process of polar code with $N=4$, where $\oplus$ represents \textbf{exclusive or} operation. The SC decoder has two types of nodes, namely $f$ node (the white one) and $g$ node (the gray one). The function of $f$ node could be described by $f(x,y)$ of Eq. (\ref{eq:n2}), which is also used in BP decoder. The function of $g$ node is defined by:
\begin{equation}\label{eq:n5}
    \begin{array}{ll}
     g(x,y,\hat{u}_{\rm{sum}})=x^{1-2\hat{u}_{\rm{sum}}}y,
	\end{array}
\end{equation}
where $\hat{u}_{\rm{sum}}=0$ or $1$. The input of SC decoder is defined by:
\begin{equation}\label{eq:n6}
    \begin{array}{ll}
     LR(y_j)= \dfrac{P(y_j | x_j = 0 )}{P(y_j | x_j = 1 )},
	\end{array}
\end{equation}
which is in the same initial value of $L(n+1,j)$ of BP decoder.

The vector $\hat{u}_1^N$ is estimated by the SC decoder sequentially. Specifically, the estimation of the bit $\hat{u}_{j}$ depends on the estimation of $\hat{u}_1^{j-1}$. Therefore, in the SC decoder, nodes in each stage are not activated simultaneously. Instead, each node is associated with a number, which is the time index indicating when the node operates. For example, in Fig. \ref{fig:n3}, the estimation of $\hat{u}_1$, $\hat{u}_2$, $\hat{u}_3$, and $\hat{u}_4$ is made at time index $2$, $3$, $5$, and $6$. The source word $\hat{u}_j$ is estimated by:

\begin{equation}\label{eq:n8}
    \begin{array}{ll}
    \hat{u}_j=\left\{
\begin{aligned}
0, &\ \ \quad \text{if} \ LR(\hat{u}_j) \geqslant 1 \ \ \text{or} \ \ j \in \mathcal{A}^c  \\
1, &\ \ \quad \text{if} \  LR(\hat{u}_j) < 1  \ \ \text{and} \ \ j \in \mathcal{A}
\end{aligned}
\right.
	\end{array}
\end{equation}

\subsection{Reformulation of Decoding Formulas} \label{sec:2.3}
To make the BP decoding and SC decoding suitable for stochastic computation, \cite{tehrani2007survey} proposed:
\begin{equation}\label{eq:n9}
    \begin{array}{ll}
    P(y_j=1)\triangleq P(y_j|x_j=1)=e^{LR(y_i)}/(1+e^{LR(y_i)}),
	\end{array}
\end{equation}
of which the value falls in range $[0,1]$, making it also suitable for molecular computation.

Therefore, for BP decoding, Eq. (\ref{eq:n0}) and Eq. (\ref{eq:n0.5}) are reformed as follows:
\begin{equation}\label{eq:n10}
    \begin{array}{ll}
     L_{n+1,j}= P(y_j=1).
	\end{array}
\end{equation}

\begin{equation}\label{eq:n11}
	R_{1,j}  = \begin{cases}
	0.5, & \text{if $j \in \mathcal{A}$,}  \\
	0, & \text{if $j \in \mathcal{A}^c$.}
	\end{cases}
\end{equation}
Eq. (\ref{eq:n2}) is reformed as:
\begin{equation}\label{eq:n12}
    \begin{aligned}
   &  F(x,y)=P_x(1-P_y)+ (1-P_x)P_y, \\
   &  G(x,y)=\dfrac{{P_x}{P_y}}{{P_x}{P_y}+(1-P_x)(1-P_y)},
	\end{aligned}
\end{equation}
where $P_x \triangleq P(x=1)$ and $P_y \triangleq P(y=1)$.

Similarly, for SC decoding, the input of the decoder is defined by $P(y_j=1)$; the output is defined by $P(\hat{u}_{j}=1)$. The function of $f$ node is also reformulated as $F(x,y)$ of Eq. (\ref{eq:n12}). As for $g$ node, Eq. (\ref{eq:n5}) is replaced by Eq. (\ref{eq:n13.5}) based on the value of $\hat{u}_{\rm{sum}}$.
\begin{equation}\label{eq:n13.5}
    \begin{aligned}
    & G(x,y,\hat{u}_{\rm{sum}}=0)=\dfrac{{P_x}{P_y}}{{P_x}{P_y}+(1-P_x)(1-P_y)},\\
   &   G(x,y,\hat{u}_{\rm{sum}}=1)=\dfrac{{(1-P_x)}{P_y}}{{(1-P_x)}{P_y}+P_x(1-P_y)}.
	\end{aligned}
\end{equation}

At last, for both BP decoding and SC decoding, the estimation of $\hat{u}_j$ is defined by:
\begin{equation}\label{eq:n15}
    \begin{array}{ll}
    \hat{u}_j=\left\{
\begin{aligned}
0, &\ \ \quad \text{if} \ P(\hat{u}_j = 1) \leqslant 0.5 \ \ \text{or} \ \ j \in \mathcal{A}^c  \\
1, &\ \ \quad \text{if} \  P(\hat{u}_j = 1) > 0.5  \ \ \text{and} \ \ j \in \mathcal{A}
\end{aligned}
\right.
	\end{array}
\end{equation}

Readers could refer to \cite{tehrani2007survey, yuan2015successive} for the derivation of the reformulated formulas.

\section{Molecular Synthesis of Decoding Formulas}\label{sec:2}

In Section \ref{sec:2.3}, three types decoding formulas need to be implemented by molecular reactions. For convenience, we rewrite them as Eq. (\ref{eq:n15.5}). Note that the three formulas are composed of two major mathematical operations, namely multiplication (e.g. $\rm{c=a \times b}$) and division (e.g. $\rm{c=a/(a+b)}$). The molecular implementation of multiplication and division is introduced in this section.
 \begin{equation}\label{eq:n15.5}
    \begin{array}{ll}
     P_z = P_x(1-P_y)+ (1-P_x)P_y, \ \ \rm{(Formula \ I)}  \\
\\
     P_z = \dfrac{{P_x}{P_y}}{{P_x}{P_y}+(1-P_x)(1-P_y)},  \ \rm{(Formula \ II)}\\
    \\
     P_z =\dfrac{{(1-P_x)}{P_y}}{{(1-P_x)}{P_y}+P_x(1-P_y)}.  \ \rm{(Formula \ III)}
	\end{array}
\end{equation}

For any variable, for example $x$,  we use the concentrations of molecular $x_1$ and $x_0$ to represent the probability distributions $P_x$ and ($1 - P_x$):
\begin{equation}\label{eq:n16}
    \begin{aligned}
     P_x &\triangleq P(x=1) = \dfrac{[x_1]}{[x_1] + [x_0]},\\
     1 - P_x &\triangleq P(x=0) = \dfrac{[x_0]}{[x_1] + [x_0]},
	\end{aligned}
\end{equation}
where $[\cdot]$ denotes the concentration.

\subsection{Probability-based Mathematical Operations}
To calculate the probability distribution of the output variable (e.g. $z$) while maintaining the probability distributions of the input variables (e.g. $x$ and $y$), we use the input variables as catalysts and molecular $S_i$ ($i=1,2...$) as the fuel of reactions. Note that all types of $S$ are in the same constant initial concentration $W$. For any variable (e.g. $x$), the sum of the concentrations of the two types of molecules are also required to be the constant number $W$ (e.g., ${[x_1]+[x_0] = W}$).

For the probability-based multiplication, e.g., $ P_z = P_x \times P_y $, reactions in Eq. (\ref{eq:n18}) proposed by \cite{xing2018dna} are adopted.
\begin{equation}\label{eq:n18}
    \begin{array}{ll}
     S_1 + x_1 \xrightarrow[]{k} T_{x_1} + x_1,\ \
     S_1 + x_0 \xrightarrow[]{k} z_0 + x_0,\\
     T_{x_1} + y_1 \xrightarrow[]{k} z_1 + y_1,\ \
     T_{x_1} + y_0 \xrightarrow[]{k} z_0 + y_0,
	\end{array}
\end{equation}

For the probability-based division, e.g., $ P_z = P_x /(P_x + P_y)$, we propose the following reactions:
\begin{equation}\label{eq:n19}
    \begin{array}{ll}
     S_1 + x_1 \xrightarrow[]{k} z_1 + x_1,\ \
     S_1 + y_1 \xrightarrow[]{k} z_0 + y_1.
	\end{array}
\end{equation}
Compared with \cite{xing2018dna}, which uses six reactions to achieve the same function $ P_z = P_x /(P_x + P_y)$, we only use two reactions here without altering the function. The ODE-based proof of Eq. (\ref{eq:n19}) is given as follows:
\begin{equation}\label{eq:n19.1}
    \begin{array}{ll}
     d[z_1]/dt = k[S_1][x_1],\ \ d[z_0]/dt = k[S_1][y_1].\\
     \\
     \Rightarrow \quad \quad \quad \quad   d[z_1]/d[z_0] = [x_1]/[y_1].
	\end{array}
\end{equation}
Given that the initial concentrations of $z_1$ and $z_0$ are both zero, and $[x_1] + [x_0] = [y_1] + [y_0] = W$, we finally have:
\begin{equation}\label{eq:n19.2}
    \begin{aligned}
    P_{z} & = \dfrac{[z_1]}{[z_1]+[z_0]} = \dfrac{[x_1]}{[x_1]+[y_1]}   \\ &=\dfrac{\frac{[x_1]}{[x_1]+[x_0]}}{\frac{[x_1]}{[x_1]+[x_0]}+\frac{[y_1]}{[y_1]+[y_0]}} = \dfrac{P_{x}}{P_{x}+P_{y}}.
	 \end{aligned}
\end{equation}

\subsection{Formula \rm{I}}
Formula $\rm{I}$ is basically the extension of probability-based multiplication, of which the molecular implementation is shown in Eq. (\ref{eq:n20}). The ODE-based simulation result of Formula $\rm{I}$ is shown in Fig. \ref{fig:n7.1}, from which can be seen that the probability calculation is essentially the process of molecule $S_1$ (fuel) being transformed into the probability distribution of output variable $z$. Since the probability distribution of input variables $x$ and $y$ are constant, their concentration curves are not shown in Fig. \ref{fig:n7.1}. The initial concentrations and the observed final concentrations of the involved molecules are listed in Table \ref{t1}, which validates Formula $\rm{I}$: $P_z = P_x (1- P_y) + (1-P_x)P_y = 0.4 \times (1-0.7) + (1-0.4) \times 0.7 =0.54$.
\begin{equation}\label{eq:n20}
    \begin{aligned}
  &   S_1 + x_1 \xrightarrow[]{k} T_{x_1} + x_1,\ \
  &   S_1 + x_0 \xrightarrow[]{k} T_{x_0} + x_0,\\
  &   T_{x_1} + y_1 \xrightarrow[]{k} {z_0} + y_1,\ \
 &    T_{x_1} + y_0 \xrightarrow[]{k} {z_1} + y_0,\\
&     T_{x_0} + y_1 \xrightarrow[]{k} {z_1} + y_1,\ \
  &   T_{x_0} + y_0 \xrightarrow[]{k} {z_0} + y_0.
	\end{aligned}
\end{equation}
\begin{table}[!htbp]
\centering
\footnotesize
\caption{Simulation results of Formula $\rm{I}$}\label{t1}
\begin{tabular}{c|c|c|c}
\Xhline{0.8pt}
 & \textbf{Molecule} & \textbf{Concentration ($\rm{Mol/L}$)} & \textbf{Probability}\\
\hline
\multirow{5}{*}{\textbf{Initial}} & $S_1$        & $10$           & $1$                    \\ \cline{2-4}
&$x_1$       & $4$            & \multirow{2}{*}{$P_x=0.4$}  \\ \cline{2-3}
&$x_0$       & $6$            &                      \\ \cline{2-4}
&$y_1$       & $7$            & \multirow{2}{*}{$P_y=0.7$}  \\ \cline{2-3}
&$y_0$       & $3$            &                      \\ \hline
\multirow{2}{*}{\textbf{Final}} &$z_1$       & $5.4$            & \multirow{2}{*}{${P_z} = 0.54$} \\ \cline{2-3}
&$z_0$       & $4.6$            &                      \\ \Xhline{0.8pt}
\end{tabular}
\end{table}

\subsection{Formula \rm{II}}
The molecular implementation of Formula $\rm{II}$ is shown in Eq. (\ref{eq:n21}), of which the first six reactions achieved multiplication and the last two reactions achieved division. Note that two types of fuel molecules ($S_1$ and $S_2$) are required in Formula $\rm{II}$ with $S_1$ for multiplication and $S_2$ for division. The simulation results are shown in Fig. \ref{fig:n7.2} and Table \ref{t2}, which validate Formula $\rm{II}$: $P_z = \frac{{P_x}{P_y}}{{P_x}{P_y}+(1-P_x)(1-P_y)}=\frac{{0.2} \times {0.6}}{{0.2}\times {0.6}+(1-0.2) \times (1-0.6)}=0.272727$.

\begin{equation}\label{eq:n21}
\resizebox{0.436\textwidth}{!}{$
    \begin{aligned}
    & S_1 + x_1 \xrightarrow[]{k} T_{x_1} + x_1,\ \ \ \ \ \ \
     S_1 + x_0 \xrightarrow[]{k} T_{x_0} + x_0,\\
   &  T_{x_1} + y_1 \xrightarrow[]{k} T_{x_1 , y_1} + y_1,\ \ \
     T_{x_1} + y_0 \xrightarrow[]{k} T_{x_1 , y_0} + y_0,\\
    & T_{x_0} + y_1 \xrightarrow[]{k} T_{x_0 , y_1} + y_1,\ \ \
     T_{x_0} + y_0 \xrightarrow[]{k} T_{x_0 , y_0} + y_0,\\
    & S_2 + T_{x_1 , y_1} \xrightarrow[]{k} z_1 +  T_{x_1 , y_1},
     S_2 + T_{x_0 , y_0} \xrightarrow[]{k} z_0 +  T_{x_0 , y_0}.
	\end{aligned}
	$}
\end{equation}

\begin{table}[!htbp]
\centering
\footnotesize
\caption{Simulation results of Formula $\rm{II}$}\label{t2}
\begin{tabular}{c|c|c|c}
\Xhline{0.8pt}
 & \textbf{Molecule} & \textbf{Concentration ($\rm{Mol/L}$)} & \textbf{Probability}\\
\hline
\multirow{6}{*}{\textbf{Initial}} & $S_1$        & $10$           & $1$                    \\ \cline{2-4}
& $S_2$        & $10$           & $1$                    \\ \cline{2-4}
&$x_1$       & $2$            & \multirow{2}{*}{$P_x=0.2$}  \\ \cline{2-3}
&$x_0$       & $8$            &                      \\ \cline{2-4}
&$y_1$       & $6$            & \multirow{2}{*}{$P_y=0.6$}  \\ \cline{2-3}
&$y_0$       & $4$            &                      \\ \hline
\multirow{6}{*}{\textbf{Final}} &$T_{x_0 , y_0}$       & $3.2$            & {$0.32$} \\ \cline{2-4}
&$T_{x_0 , y_1}$       & $4.8$            &              {$0.48$}         \\ \cline{2-4}
&$T_{x_1 , y_0}$       & $0.8$            & {$0.08$} \\ \cline{2-4}
&$T_{x_1 , y_1}$       & $1.2$            &           {$0.12$}           \\ \cline{2-4}
&$z_1$       & $2.72727$            &  \multirow{2}{*}{${P_z} = 0.272727$} \\ \cline{2-3}
&$z_0$       & $7.27272$            &                     \\ \Xhline{0.8pt}
\end{tabular}
\begin{tablenotes}
        \footnotesize
       \item[1] \scriptsize {The probability of $T_{x_i,y_j}$ is defined as: $\frac{[T_{x_i,y_j}]}{[T_{x_0,y_0}]+[T_{x_0,y_1}]+[T_{x_1,y_0}]+[T_{x_1,y_1}]}$.}
      \end{tablenotes}
\end{table}

\subsection{Formula \rm{III}}
\begin{figure}[ht]
\centering
\subfigure[ODE-based simulation results of Formula $\rm{I}$.]{
\label{fig:n7.1}
\includegraphics[width=0.7\linewidth]{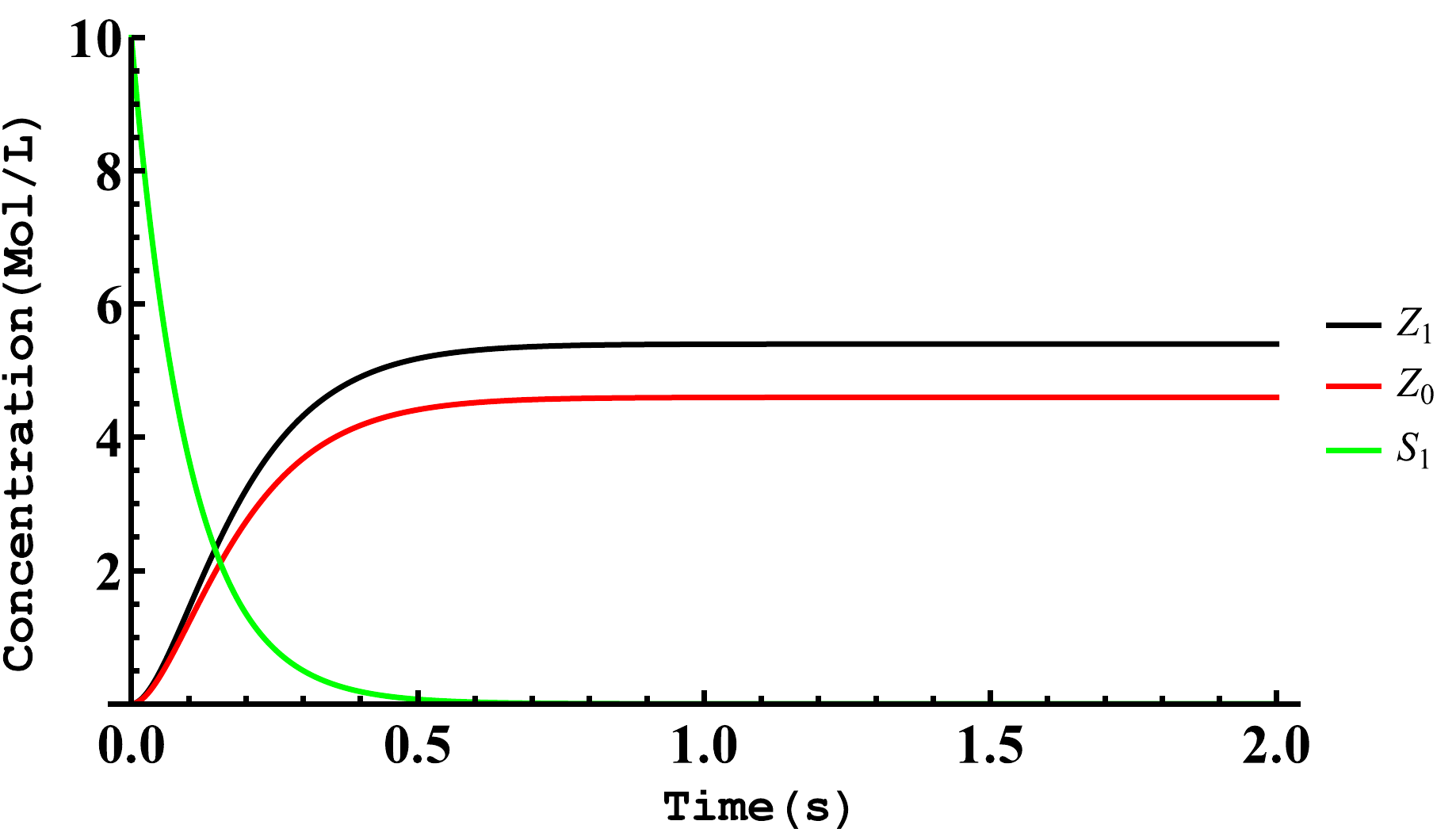}}
\subfigure[ODE-based simulation results of Formula $\rm{II}$.]{
\label{fig:n7.2}
\includegraphics[width=0.7\linewidth]{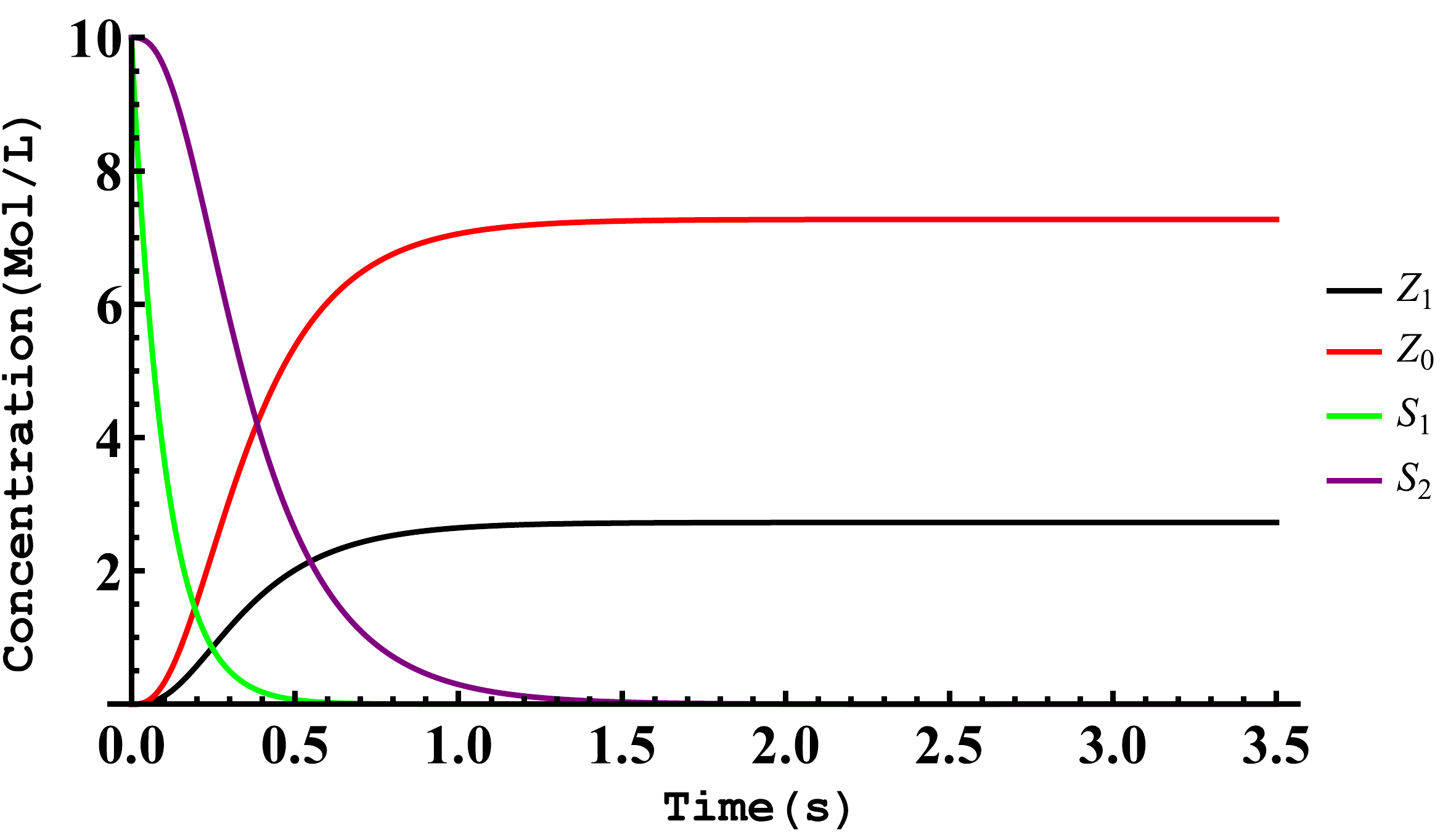}}
\subfigure[ODE-based simulation results of Formula $\rm{III}$.]{
\label{fig:n7.3}
\includegraphics[width=0.7\linewidth]{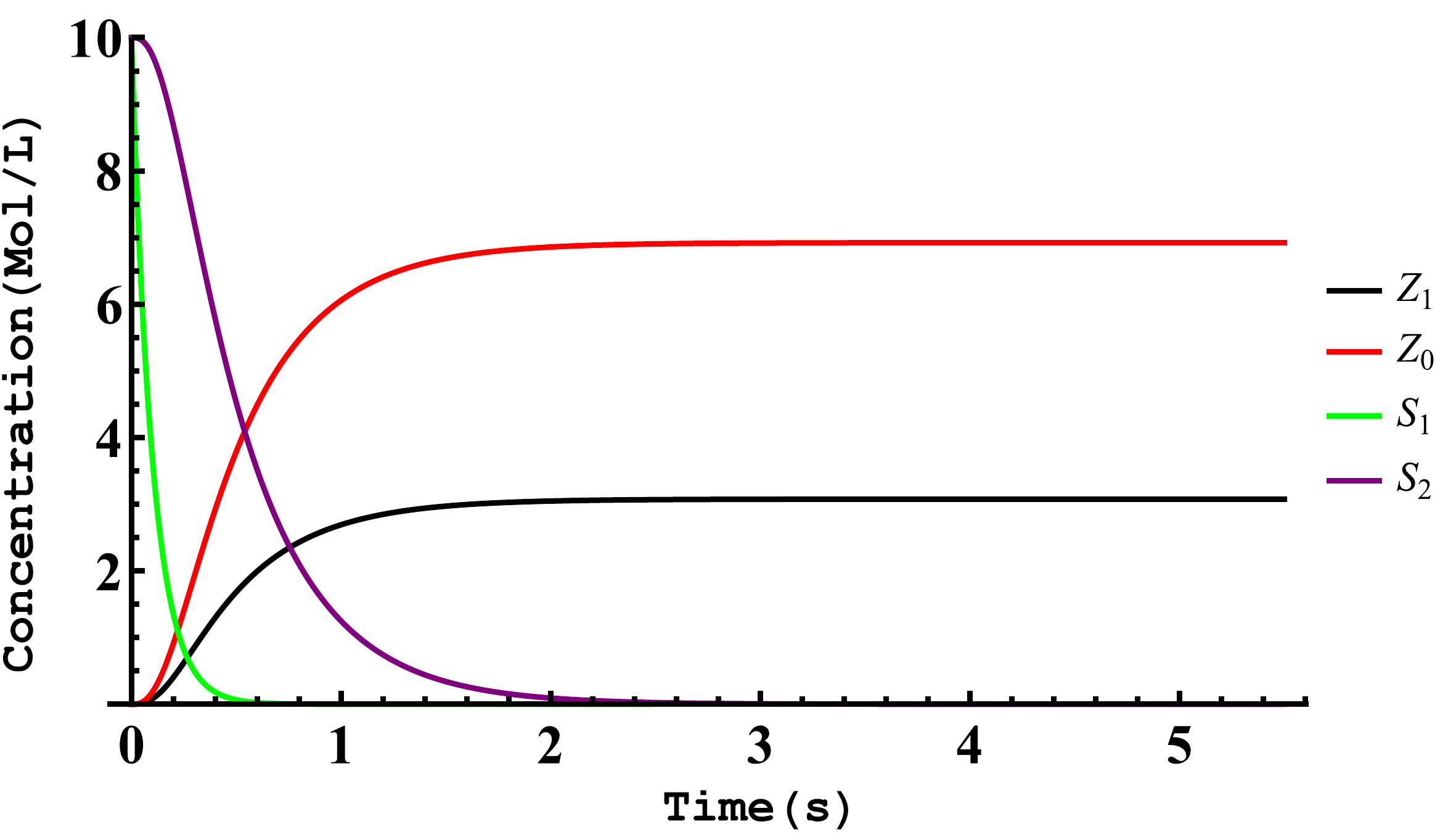}}
\caption{Simulation results of polar decoding formulas.}\label{fig:n777}
\end{figure}

The molecular implementation of Formula $\rm{III}$, shown in Eq. (\ref{eq:n22}), is similar to that of Formula $\rm{II}$, since they are in the same structure. In Eq. (\ref{eq:n22}), the first six reactions for multiplication are the same as that of Eq. (\ref{eq:n21}). The difference between Eq. (\ref{eq:n22}) and  Eq. (\ref{eq:n21}) is that the reactants in the last two reactions for division are different. The simulation results are represented by Fig. \ref{fig:n7.3} and Table \ref{t3}, which validate Formula $\rm{III}$: $P_z =\frac{{(1-P_x)}{P_y}}{{(1-P_x)}{P_y}+P_x(1-P_y)}=\frac{{(1-0.9)} \times {0.8}}{{(1-0.9)} \times {0.8}+0.9 \times (1-0.8)}=0.307692$.
\begin{equation}\label{eq:n22}
\resizebox{0.436\textwidth}{!}{$
    \begin{aligned}
   &  S_1 + x_1 \xrightarrow[]{k} T_{x_1} + x_1,\ \ \ \ \ \ \
     S_1 + x_0 \xrightarrow[]{k} T_{x_0} + x_0,\\
   &  T_{x_1} + y_1 \xrightarrow[]{k} T_{x_1 , y_1} + y_1,\ \ \
     T_{x_1} + y_0 \xrightarrow[]{k} T_{x_1 , y_0} + y_0,\\
   &  T_{x_0} + y_1 \xrightarrow[]{k} T_{x_0 , y_1} + y_1,\ \ \
     T_{x_0} + y_0 \xrightarrow[]{k} T_{x_0 , y_0} + y_0,\\
  &   S_2 + T_{x_0 , y_1} \xrightarrow[]{k} z_1 +  T_{x_0 , y_1},
     S_2 + T_{x_1 , y_0} \xrightarrow[]{k} z_0 +  T_{x_1 , y_0}.
	\end{aligned}
	$}
\end{equation}
\begin{table}[!htbp]
\centering
\footnotesize
\caption{Simulation results of Formula $\rm{III}$}\label{t3}
\begin{tabular}{c|c|c|c}
\Xhline{0.8pt}
 & \textbf{Molecule} & \textbf{Concentration ($\rm{Mol/L}$)} & \textbf{Probability}\\
\hline
\multirow{6}{*}{\textbf{Initial}} & $S_1$        & $10$           & $1$                    \\ \cline{2-4}
& $S_2$        & $10$           & $1$                    \\ \cline{2-4}
&$x_1$       & $9$            & \multirow{2}{*}{$P_x=0.9$}  \\ \cline{2-3}
&$x_0$       & $1$            &                      \\ \cline{2-4}
&$y_1$       & $8$            & \multirow{2}{*}{$P_y=0.8$}  \\ \cline{2-3}
&$y_0$       & $2$            &                      \\ \hline
\multirow{6}{*}{\textbf{Final}} &$T_{x_0 , y_0}$       & $0.2$            & {$0.02$} \\ \cline{2-4}
&$T_{x_0 , y_1}$       & $0.8$            &              {$0.08$}         \\ \cline{2-4}
&$T_{x_1 , y_0}$       & $1.8$            & {$0.18$} \\ \cline{2-4}
&$T_{x_1 , y_1}$       & $7.2$            &           {$0.72$}           \\ \cline{2-4}
&$z_1$       & $3.07692$            &  \multirow{2}{*}{${P_z} = 0.307692$} \\ \cline{2-3}
&$z_0$       & $6.92308$            &                     \\ \Xhline{0.8pt}
\end{tabular}
\begin{tablenotes}
        \footnotesize
       \item[1] \scriptsize {The probability of $T_{x_i,y_j}$ is defined as: $\frac{[T_{x_i,y_j}]}{[T_{x_0,y_0}]+[T_{x_0,y_1}]+[T_{x_1,y_0}]+[T_{x_1,y_1}]}$.}
      \end{tablenotes}
\end{table}

\section{Implementation and Simulations}\label{sec:3}
In this section, we propose the implementation and simulation results of polar BP decoder and SC decoder. The concentrations of all types of $S$ are set to be $W=10$ $\rm{Mol/L}$. All reactions share the same rate constant $k=1\rm{M}^{-1} \rm{s}^{-1}$. The sum of concentrations of two types of molecules of any variable is  $W=10$ $\rm{Mol/L}$. CRNs are solved numerically with \textbf{Mathematica} based on the platform provided by \cite{Simul}. For convenience, we refer to symbol $\cdot[h=r]$ as the molecule indicating $h=r$ ($h$ could be any variable and $r$ is its value).

\begin{table*}[!t]
\centering
\caption{Simulation Results of BP decoder with $N=8$ During $6$ Iterations}
\label{t6}
\begin{tabular}{c||c|c|c|c|c|c|c|c}
\Xhline{0.8pt}
\textbf{Iteration}  & $L_{1,1}$  & $L_{1,2}$  & $L_{1,3}$  & $L_{1,4}$  & $L_{1,5}$ & $L_{1,6}$  & $L_{1,7}$  & $L_{1,8}$    \\ \hline

${\textbf{1}}$ & $0.7$ & $0.57764$ & $0.5$ & $0.549407$ & $0.6$ & $0.205882$ & $0.931034$ & $0.272727$   \\ \hline
${\textbf{2}}$ & $0.463697$ & $0.538628$ & $0.445218$ & $0.664612$ & $0.467507$ & $0.251153$ & $0.830331$ & $0.208378$   \\ \hline
${\textbf{3}}$ & $0.524419$ & $0.535672$ & $0.468193$ & $0.668301$ & $0.455842$ & $0.219076$ & $0.808274$ & $0.146418$   \\ \hline
${\textbf{4}}$ & $0.520739$ & $0.534052$ & $0.468556$ & $0.659226$ & $0.460047$ & $0.238122$ & $0.811173$ & $0.163673$  \\ \hline
${\textbf{5}}$ & $0.518814$ & $0.534172$ & $0.467884$ & $0.661556$ & $0.460169$ & $0.238652$ & $0.811952$ & $0.164472$   \\ \hline
${\textbf{6}}$ & $0.518981$ & $0.534226$ & $0.467887$ & $0.661821$ & $0.460054$ & $0.238116$& $0.811858$ & $0.163998$   \\ \Xhline{0.8pt}
\end{tabular}
\end{table*}

\subsection{Implementation of BP Decoder}
For polar BP decoder with $n=\log_{2}N$ stages, one iteration of message passing refers to the activation of BCBs from stage $n$ to stage $1$ for the update on messages $L$, and the activation of BCBs from stage $1$ to stage $n$ for the update on messages $R$. Note that $L_{n+1,j}$ and $R_{1,j}$ are not required to be updated because they are determined by the channel; $R_{n+1,j}$ are not required to be updated because they are not involved in the calculation of other messages ($j=1,2,...,N$). For instance, the order of messages updated in Fig. \ref{fig:n1} in one iteration is: $L_{3,j}$, $L_{2,j}$, $L_{1,j}$, $R_{2,j}$, and  $R_{3,j}$. Messages $L$ (or $R$) in the same stage are updated simultaneously.

According to Eq. (\ref{eq:n1}), the updating of one message needs both Formula $\rm{I}$ and Formula $\rm{II}$, thus three types of $S$ are required. Here we use symbol {\small\text{$S_{L_{i,j}}^t$}} (or {\small\text{$S_{R_{i,j}}^t$}}) to indicate the three types of $S$ that are needed in the updating of message $L_{i,j}$ (or $R_{i,j}$) in $t$-th iteration. During BP decoding, molecules of which the concentrations represent the initial values of $L$ and $R$ are stored in the solution at first. The BP decoder is activated and controlled by injecting {\small\text{$S_{L_{i,j}}^t$}} and {\small\text{$S_{R_{i,j}}^t$}} into the solution to activate BCBs in the sequence mentioned above. The outputs of BCBs in stage $i$ will be the inputs of BCBs in stage ($i-1$) or ($i+1$). Given that the premise of successful implementation of decoding formulas is that the sum of concentrations of two types of molecules for the input variable is constant (e.g. $[x_1] + [x_0] = W$), the interval between injections of {\small\text{$S_{L_{i,j}}^t$}} (or {\small\text{$S_{R_{i,j}}^t$}}) for the activation of adjacent stages should be long enough, so that the reactions in the former stage could go to completion, resulting in the constant concentrations of input variables for the next stage.

\begin{figure}[ht]
\centering
\includegraphics[width=8.5cm]{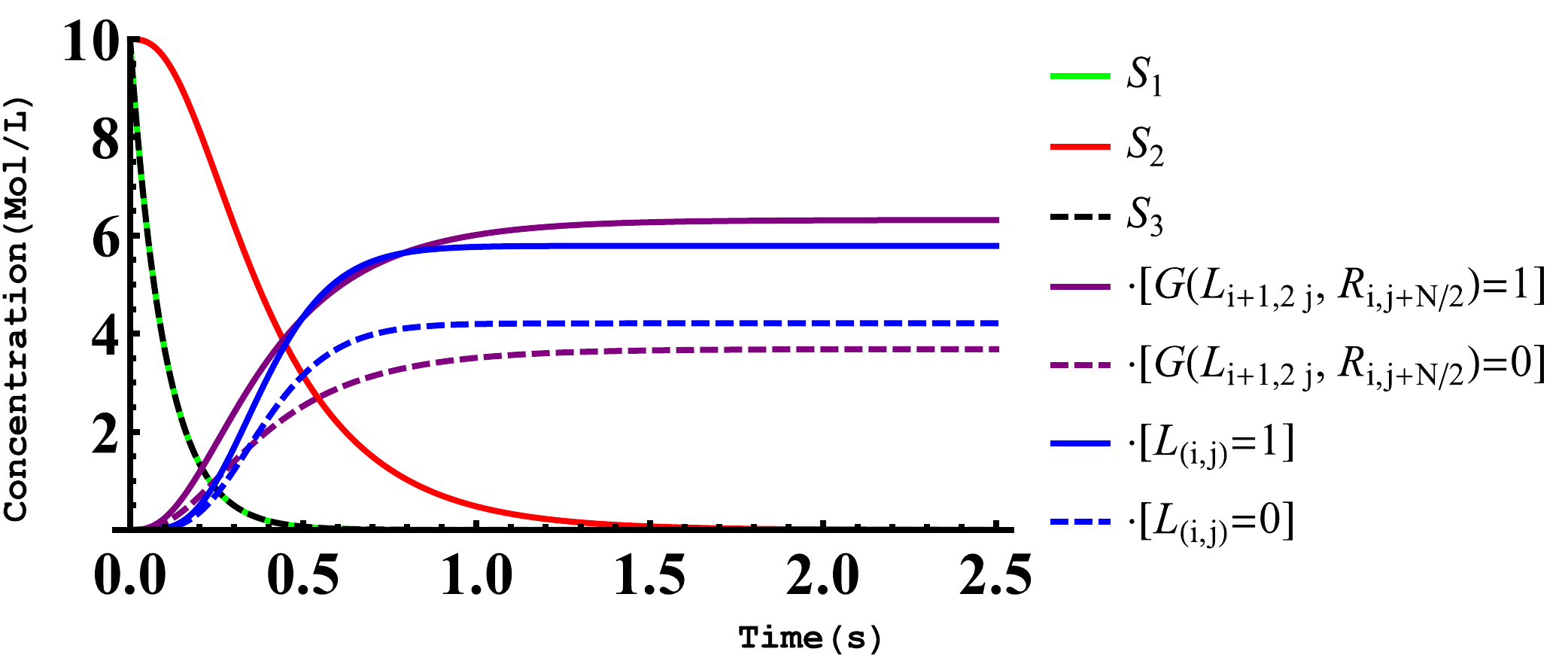}
\caption{Simulation results of $L_{i,j}=f(L_{i+1,2j-1},g(L_{i+1,2j},R_{i,j+N/2}))$.}\label{fig:n6}
\end{figure}

Fig. \ref{fig:n6} illustrates the ODE-based simulation results of $L_{i,j}=f(L_{i+1,2j-1},g(L_{i+1,2j},R_{i,j+N/2}))$, where $L_{i+1,2j-1}=0.2$, $L_{i+1,2j}=0.8$ and $R_{i,j+N/2}=0.3$. The final concentrations of molecules $\cdot[g(L_{i+1,2j},R_{i,j+N/2})=1]$ and $\cdot[L_{i,j}=1]$ show that the $g(L_{i+1,2j},R_{i,j+N/2})=0.631579$ and $L_{i,j}=0.578947$, which correspond with the theoretical results.

The simulation results of BP decoder with $N=8$ during $6$ iterations are listed in Table \ref{t6}. The inputs of the BP decoder are $L_{4,1} = 0.7$, $L_{4,2} = 0.4$, $L_{4,3} = 0.3$, $L_{4,4} = 0.7$, $L_{4,5} = 0.6$, $L_{4,6} = 0.1$, $L_{4,7} = 0.9$, $L_{4,8} = 0.2$. $u_1$, $u_4$, $u_6$, $u_7$, and $u_8$ are chosen as the information bits, so that we have the following initializations: $R_{1,1} = R_{1,4} = R_{1,6} = R_{1,7} = R_{1,8} = 0.5$ and $R_{1,2} = R_{1,3} = R_{1,5} = 0$. The interval between injections of {\small\text{$S_{L_{i,j}}^t$}} (or {\small\text{$S_{R_{i,j}}^t$}}) for adjacent stages is set to be $10s$. The values of $L_{1,j}$ $(j=1,2,...8)$ in Table \ref{t6} are obtained by recording the concentrations of molecules $\cdot[L_{1,j}=1]$ in the CRNs and are exactly the same with the theoretical results. Note that the input of the BP decoder and the information bits of the polar code are set randomly, because we only aim to demonstrate the feasibility of the molecular implementation of BP decoder rather than test or improve its decoding performance.

\subsection{Implementation of SC Decoder}

Similar to BP decoder, each $g$ node or $f$ node in SC decoder needs three types of $S$ to achieve its function. Since each node is associated with a time index for activation, the sequence of injections of $S$ for each node also obeys the the time index. During SC decoding, the estimation of $\hat{u}_{j}$ should be made immediately after $LR(\hat{u}_{j})$ is figured out since the calculation of $LR(\hat{u}_{j+i})$ depends on the value of $\hat{u}_{j}$.

According Eq. (\ref{eq:n15}), $\hat{u}_{j}$ is estimated to be $0$ if it is the frozen bit. Such estimation could be achieved by setting the concentration of $\cdot[LR(\hat{u}_{j})=0]$ to be $W$ and the concentration of $\cdot[LR(\hat{u}_{j})=1]$ to be $0$. When $\hat{u}_{j}$ is the information bit, it is estimated to be $0$ if $P(\hat{u}_{j}=1) < P(\hat{u}_{j}=0)$ or to be $1$ if $P(\hat{u}_{j}=1) \geqslant P(\hat{u}_{j}=0)$. This could be implemented by applying a consensus network proposed in \cite{chen2013programmable} to  $\cdot[LR(\hat{u}_{j})=0]$ and $\cdot[LR(\hat{u}_{j})=1]$. A consensus network (e.g. Eq. (\ref{eq:n23})) operates on two types of molecules (e.g. \(h_1\) and \(h_0\)); it converts the molecule in lower concentration into the other molecule in higher concentration. Therefore, the consensus network is able to decide the value of $\hat{u}_{j}$ for both polar BP decoder and SC decoder.
\begin{equation}\label{eq:n23}
    \begin{aligned}
     h_0 + h_1 \xrightarrow[]{k} 2B, \ \
     h_1 + B \xrightarrow[]{k} 2{h_1}, \ \
      h_0 + B \xrightarrow[]{k} 2{h_0}.
      \end{aligned}
\end{equation}

After the estimation of $\hat{u}_{j}$, the specific function of relevant $g$ nodes could be determined. For example, for the $g$ node labelled with time index $3$ in Fig. \ref{fig:n3}, its function is defined by Formula $\rm{II}$ if $\hat{u}_{1} =0$; or its function is defined by Formula $\rm{III}$ if $\hat{u}_{1} =1$. The choice between Formula $\rm{II}$ and Formula $\rm{III}$ for a $g$ node is achieved by using $\cdot[LR(\hat{u}_{j})=0]$ or $\cdot[LR(\hat{u}_{j})=1]$ to convert the injected $S$ into the one that could activate Formula $\rm{II}$ or Formula $\rm{III}$. Therefore, in SC decoder, the injected $S$ for $g$ nodes could not activate any computation unless being transformed. For example, for the $g$ node labelled with time index $3$ in Fig. \ref{fig:n3}, assume $S_1$ is the injected molecule, $S_2$ or  $S_3$ could activate Formula $\rm{II}$ or Formula $\rm{III}$, respectively. The choice between Formula $\rm{II}$ and Formula $\rm{III}$ is achieved by:
\begin{equation}\label{eq:n24}
    \begin{aligned}
     S_1 + \cdot[LR(\hat{u}_{1}) =0] \xrightarrow[]{k} S_2 + \cdot[LR(\hat{u}_{1}) =0], \\
     S_1 + \cdot[LR(\hat{u}_{1}) =1] \xrightarrow[]{k} S_3 + \cdot[LR(\hat{u}_{1}) =1].
      \end{aligned}
\end{equation}

For $g$ nodes that take $\hat{u}_{i} \oplus ... \oplus \hat{u}_{j}$ as input, intermediate products are introduced in the transformation from the injected $S$ to the functional $S$. For example, for the $g$ node labelled with time index $4$ in Fig. \ref{fig:n3}, assume $S_4$ is the injected molecule, $S_5$ or  $S_6$ could activate Formula $\rm{II}$ or Formula $\rm{III}$, respectively. The choice between Formula $\rm{II}$ and Formula $\rm{III}$ is achieved by:
\begin{equation}\label{eq:n25}
    \begin{aligned}
     S_4 + \cdot[LR(\hat{u}_{1}) =0] \xrightarrow[]{k} I_1 + \cdot[LR(\hat{u}_{1}) =0], \\
     S_4 + \cdot[LR(\hat{u}_{1}) =1] \xrightarrow[]{k} I_2 + \cdot[LR(\hat{u}_{1}) =1], \\
     I_1 + \cdot[LR(\hat{u}_{2}) =0] \xrightarrow[]{k} S_5 + \cdot[LR(\hat{u}_{1}) =0], \\
     I_1 + \cdot[LR(\hat{u}_{2}) =1] \xrightarrow[]{k} S_6 + \cdot[LR(\hat{u}_{1}) =1],\\
     I_2 + \cdot[LR(\hat{u}_{2}) =0] \xrightarrow[]{k} S_6 + \cdot[LR(\hat{u}_{1}) =0], \\
     I_2 + \cdot[LR(\hat{u}_{2}) =1] \xrightarrow[]{k} S_5 + \cdot[LR(\hat{u}_{1}) =1],
      \end{aligned}
\end{equation}
where $I_n$ ($n=1,2$) are the intermediate products.

\begin{figure}[ht]
\centering
\includegraphics[width=7.5cm]{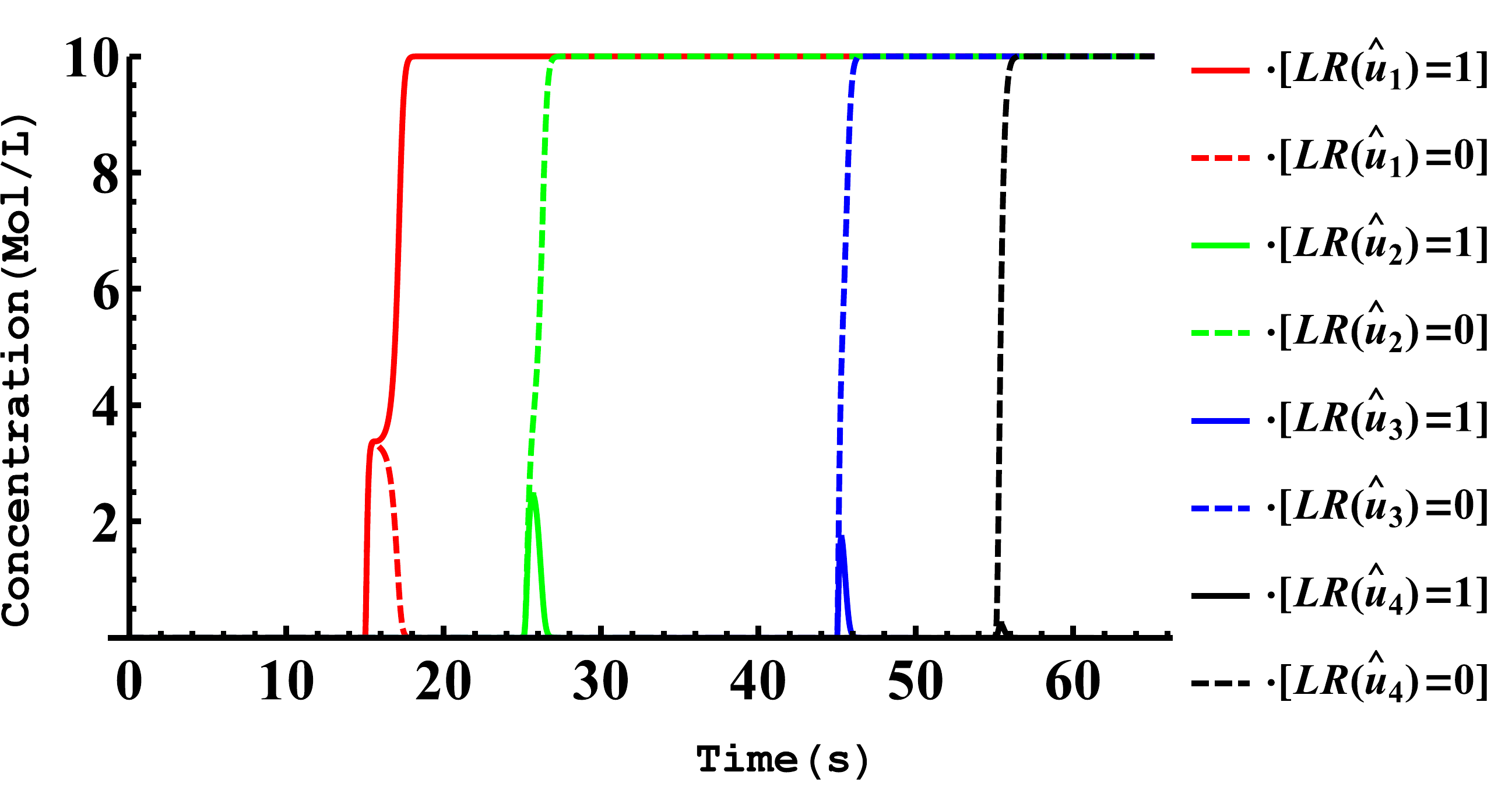}
\caption{ODE-based simulation results of SC decoder with $N=4$ .}\label{fig:n7}
\end{figure}

Fig. \ref{fig:n7} illustrates the simulation results of SC decoder with $N=4$. The inputs are $LR(y_1)=0.7$, $LR(y_2)=0.48$, $LR(y_3)=0.4$, and $LR(y_4)=0.2$. The outputs come out in sequence: $LR(\hat{u}_1)=1$, $LR(\hat{u}_2)=0$, $LR(\hat{u}_3)=0$, and $LR(\hat{u}_4)=0$. So the estimations of source word are: $\hat{u}_{1} =1$, $\hat{u}_{2} = \hat{u}_{3} =\hat{u}_{4} =0$, which are exactly the same with the theoretical results. Note that the concentration of $\cdot[LR(\hat{u}_1) =0]$ (red dashed curve) dropped to zero because the consensus network transformed $\cdot[LR(\hat{u}_1) =0]$ into $\cdot[LR(\hat{u}_1) =1]$, which was in higher concentration.

\section{Remarks}\label{sec:4}

This paper proposes a method of synthesizing polar BP decoder and SC decoder with molecular reactions. Both types of decoders are controlled by injecting multiple types of $S$ to trigger Formula $\rm{I}$, $\rm{I}$ or $\rm{III}$. The molecular implementation of the feedback part ($\hat{u}_{i} \oplus ... \oplus \hat{u}_{j}$) of SC decoder is also given. In \citep{xing2018dna}, Formula $\rm{I}$ and $\rm{II}$ are also implemented with $14$ and $19$ molecular reactions , respectively. However, our method only needs $6$ and $8$ reaction for Formula $\rm{I}$ and $\rm{II}$, respectively.

Strictly speaking, there are three types of reaction involved in our design. The first type is $A + B \xrightarrow[]{k} C + B$, which is used in the reactions implementing decoding formulas. The rest of the reaction types are  $A + B \xrightarrow[]{k} 2C $ and $A + B \xrightarrow[]{k} 2B$ which are used in the consensus networks. The three types of reactions have been physically implemented with DNA strand displacement reactions by \citep{chen2013programmable}, making our design promising for the application in nanoscale devices in the future.

The numerical simulations of polar decoders yield exact results because the essence of CRN-based BP decoder and SC decoder is ODEs. Although the simulation results have validated the feasibility and robustness of our method and all the reactions involved have been achieved by DNA reactions, there is prerequisite for the DNA-based implementation of polar decoders: 1) the rate constants of all reactions should be the same; 2) the sum of concentrations of two types of molecules of any variable should be constant. Our primary contribution is the ODE-based model of polar BP decoder and SC decoder, which needs fewer reactions for the decoding formulas compared with \citep{xing2018dna} and only employs reaction types that are DNA-implementation-friendly.

\small
\bibliographystyle{IEEEtran}
%% argument is your BibTeX string definitions and bibliography database(s)
\bibliography{IEEEabrv,mybib}

% that's all folks
\end{document}